\def \SAIT #1 #2 {{\em Mem.\ Soc.\ Astron.\ It.\/} {\bf #1}, #2}
\def \MESS #1 #2 {{\em The Messenger\/} {\bf #1}, #2}
\def \ASTRNACH #1 #2 {{\em Astron. Nach.\/} {\bf #1}, #2}
\def \AAP #1 #2 {{\em Astron. Astrophys.\/} {\bf #1}, #2}
\def \AAL #1 #2 {{\em Astron. Astrophys. Lett.\/} {\bf #1}, L#2}
\def \AAR #1 #2 {{\em Astron. Astrophys. Rev.\/} {\bf #1}, #2}
\def \AAS #1 #2 {{\em Astron. Astrophys. Suppl. Ser.\/} {\bf #1}, #2}
\def \AJ #1 #2 {{\em Astron. J.\/} {\bf #1}, #2}
\def \ANNREV #1 #2 {{\em Ann. Rev. Astron. Astrophys.\/} {\bf #1}, #2}
\def \APJ #1 #2 {{\em Astrophys. J.\/} {\bf #1}, #2}
\def \APJL #1 #2 {{\em Astrophys. J. Lett.\/} {\bf #1}, L#2}
\def \APJS #1 #2 {{\em Astrophys. J. Suppl.\/} {\bf #1}, #2}
\def \APSS #1 #2 {{\em Astrophys. Space Sci.\/} {\bf #1}, #2}
\def \ASR #1 #2 {{\em Adv. Space Res.\/} {\bf #1}, #2}
\def \BAIC #1 #2 {{\em Bull. Astron. Inst. Czechosl.\/} {\bf #1}, #2}
\def \JSQRT #1 #2 {{\em J. Quant. Spectrosc. Radiat. Transfer\/} {\bf #1}, #2}
\def \MN #1 #2 {{\em Mon. Not. R. Astr. Soc.\/} {\bf #1}, #2}
\def \MEM #1 #2 {{\em Mem. R. Astr. Soc.\/} {\bf #1}, #2}
\def \PLR #1 #2 {{\em Phys. Lett. Rev.\/} {\bf #1}, #2}
\def \PASJ #1 #2 {{\em Publ. Astron. Soc. Japan\/} {\bf #1}, #2}
\def \PASP #1 #2 {{\em Publ. Astr. Soc. Pacific\/} {\bf #1}, #2}
\def \NAT #1 #2 {{\em Nature\/} {\bf #1}, #2}
\def\HI{\hbox{H~$\scriptstyle\rm I\ $}}
\def\HII{\hbox{H~$\scriptstyle\rm II\ $}}
\def\nHI{{\rm HI}}
\def\nH{{\rm H}}
\def\nHII{{\rm HII}}

\def\HeII{\hbox{He~$\scriptstyle\rm II\ $}}

\def\HeIII{\hbox{He~$\scriptstyle\rm III\ $}}

\def\kmsmpc{\,{\rm km\,s^{-1}\,Mpc^{-1}}}
\def\cmm{\,{\rm cm^{-2}}}

\def\ndotunits{\,{\rm phot\,s^{-1}\,Mpc^{-3}}}

\def\sfrd{\,{\rm M_\odot\,yr^{-1}\,Mpc^{-3}}}
\def\sfr{\,{\rm M_\odot\,yr^{-1}}}

\def\Lya{Ly$\alpha\ $}
\def\etal{{et al.\ }}

\def\spose#1{\hbox to 0pt{#1\hss}}
\def\lta{\mathrel{\spose{\lower 3pt\hbox{$\mathchar"218$}}
     \raise 2.0pt\hbox{$\mathchar"13C$}}}
\def\gta{\mathrel{\spose{\lower 3pt\hbox{$\mathchar"218$}}
     \raise 2.0pt\hbox{$\mathchar"13E$}}}

\documentstyle{memsait}
\input epsf.sty
\begin{opening}
\title{WHAT KEEPS THE UNIVERSE IONIZED AT $Z\approx 5$?}
\author{PIERO MADAU}
\institute{Space Telescope Science Institute, Baltimore, MD 21218, USA}
\date{} 
\end{opening}

\begin{document}

\oddpagefooter{}{}{} 
\evenpagefooter{}{}{} 
\ 
\bigskip

\begin{abstract}
The history of the transition from a neutral intergalactic medium to one
that is almost fully ionized can reveal the character of cosmological ionizing
sources. In this talk I will discuss the implications for rival reionization 
scenarios of the rapid decline observed in the space
density of quasars and star-forming galaxies at redshifts $z\gta
3$. The hydrogen component in a highly inhomogeneous universe is completely 
reionized when the number of ionizing photons emitted in one recombination 
time equals the mean number of hydrogen atoms. At $z\sim 5$, the local character of
the UV metagalactic flux allows one to define a {\it critical} emission rate of
hydrogen-ionizing photons per unit comoving volume, ${\dot{\cal N}}_{\rm
ion}=10^{51.5\pm 0.3}\ndotunits$. Models based on photoionization by bright
QSOs and/or young galaxies with star formation rates in excess of $0.3-1\sfr$ 
appear to fail to provide the required number of hydrogen-ionizing
photons at these redshifts by large factors. If stellar sources are 
responsible for keeping the universe ionized at $z\approx 5$, the rate of star 
formation per unit comoving volume at this epoch must be comparable or greater
than observed at $z\approx 3$.
\end{abstract}

\section{Introduction}

The existence of a filamentary, low-density intergalactic medium (IGM) which
contains the bulk of the hydrogen and helium in the universe is predicted as a
product of primordial nucleosynthesis and
of hierarchical models of gravitational instability with ``cold dark matter''
(CDM) (Cen \etal 1994; Zhang \etal 1995; Hernquist \etal 1996).
The application of the Gunn-Peterson constraint on the amount of smoothly
distributed neutral material along the line of sight to distant objects 
requires the hydrogen component of the diffuse IGM to have been highly ionized 
by $z\approx 5$ (Schneider \etal 1991), and the helium component by $z\approx 
2.5$ (Davidsen \etal 1996). It thus appears that substantial sources of ultraviolet
photons were present at $z\gta 5$, perhaps low-luminosity quasars or a first 
generation of stars in virialized dark matter halos with
$T_{\rm vir}\gta 10^4\,$K (Couchman \& Rees 1986; Ostriker \& Gnedin 1996; 
Haiman \& Loeb 1997; Miralda-Escud\`e \& Rees 1997). Early star
formation provides a possible explanation for the widespread existence of heavy
elements in the IGM (Cowie \etal 1995), while reionization by QSOs may
produce a detectable signal in the radio extragalactic background at meter
wavelengths (Madau \etal 1997). Establishing the character of
cosmological ionizing sources is an efficient way to constrain competing
models for structure formation in the universe, and to study the collapse and
cooling of small mass objects at early epochs. While the nature, spectrum, and 
intensity of the background UV flux which is responsible for maintaining the 
intergalactic gas and the Ly$\alpha$ clouds in a highly ionized state at 
$z\lta 3$ has been the subject of much debate in the last decade, it is only in 
the past few years that new 
observations have provided reliable information on the presence and physical
properties of the sources and sinks (due to continuum opacities) of UV radiation 
in the interval $3\lta z\lta 5$. 

In this talk I will focus on the candidate sources of photoionization at 
early times and on the time-dependent reionization problem, i.e. on the history
of the transition from a neutral IGM to one that is almost fully ionized. The 
starting point of this study can be found perhaps in the simple realization
that the {\it breakthrough epoch} (when all radiation sources can see 
each other in the Lyman continuum) occurs much later in the universe than 
the {\it overlap epoch} (when individual ionized zones become simply 
connected and every point in space is exposed to ionizing radiation), and that
at high redshifts the ionization equilibrium is actually determined by
the {\it instantaneous} UV production rate. 
In the following I will adopt an Einstein-de Sitter universe ($q_0=0.5$) 
with $H_0= 50h_{50}\,\kmsmpc$.

\section{The Universe After Complete Overlapping}

The complete reionization of the universe manifests itself in the absence of a
Gunn-Peterson absorption trough in galaxies and quasars at $z\lta 5$. In
the presence of a uniform medium of \HI density $n_\nHI(z)$ along the 
path to a distant object, the optical depth associated with resonant \Lya 
scattering at $\lambda_\alpha(1+z)$ is 
\begin{equation}
\tau_{\rm GP}(z)={\pi e^2 f_\alpha \lambda_\alpha n_\nHI(z)\over m_ec H}=
8.3\times 10^{10} h_{50}^{-1} {n_\nHI(z)\over (1+z)^{3/2}}, \label{eq:gp} 
\end{equation}
where $H$ is the Hubble constant, $f_\alpha$ is the oscillator strength of 
the transition, and all other symbols have their usual meaning. The strongest 
limit on the amount of diffuse intergalactic neutral hydrogen at $z=4.3$ is 
provided by a direct estimate of the quasar Q1202--0725 flux in regions of 
the spectrum where lines are absent (Giallongo \etal 1994): the method yields
$\tau_{\rm GP}\le 0.02\pm 0.03$. Even if 99\% of all the cosmic 
baryons fragment at these epochs into structures that can be identified 
with discrete absorption lines, with only 1\% remaining in a smoothly
distributed component (cf. Zhang \etal 1998), these measurements imply a diffuse 
IGM which is ionized to better than 1 part in $10^4$.  

It is useful to start the discussion with the last stage of the 
reionization process,
when individual ionized zones have overlapped, the reionization of hydrogen 
(and helium) in the universe has been completed, and the IGM has been exposed
everywhere to Lyman continuum photons. An often overlooked point is worth 
remarking here, namely the fact that even if the bulk of the baryons in the 
universe are fairly well ionized at $z\lta 5$, the neutral hydrogen still 
present in the numerous Ly$\alpha$ forest clouds and the rarer Lyman-limit 
systems (LLS) along the line of sight significantly attenuates the ionizing 
flux from cosmologically distant sources. In other words, while the complete
overlapping of I-fronts occurs at $z_{\rm ov}>5$, the universe does not 
become optically thin to Lyman continuum photons until much later, at a 
``breakthrough'' redshift $z_{\rm br}\approx 1.6$. It is only 
after breakthrough that any point in space at any redshift $z<z_{\rm br}$ 
will be exposed to ionizing photons from all radiation sources out to $z_{\rm
br}$.

\subsection{Cosmological Radiative Transfer}

The radiative transfer equation in cosmology describes the evolution in time of
the specific intensity $J$ of a diffuse radiation field: 
\begin{equation}
\left({\partial \over \partial t}-\nu {\dot a \over a} {\partial
\over \partial \nu}\right)J=-3{\dot a \over a}J-c\kappa J + 
{c\over 4\pi}\epsilon, \label{eq:rad} 
\end{equation}
where $a$ is the cosmological scale parameter, $c$ the speed of the light,
$\kappa$ the continuum absorption coefficient per unit length along the line of
sight, and $\epsilon$ is the proper space-averaged volume emissivity. The mean
(averaged over all lines of sight) specific intensity of the radiation
background at the observed frequency $\nu_o$, as seen by an observer at
redshift $z_o$, is then 
\begin{equation}
J(\nu_o,z_o)={1\over 4\pi}\int_{z_o}^{\infty}\, dz\, {dl \over dz} 
{(1+z_o)^3 \over (1+z)^3} \epsilon(\nu,z)e^{-\tau_{\rm eff}}, \label{eq:Jnu} 
\end{equation}
where $\nu=\nu_o(1+z)/(1+z_o)$, and $dl/dz$ is the line element in a Friedmann
cosmology. The effective optical depth $\tau_{\rm eff}$ due to
discrete absorption systems is defined, for Poisson-distributed clouds, as
\begin{equation}
\tau_{\rm eff}(\nu_o,z_o,z)=\int_{z_o}^z\, dz'\int_0^{\infty}\, dN_\nHI\,
{\partial^2 N \over \partial N_\nHI \partial z'} (1-e^{-\tau}) \label{eq:tau} 
\end{equation}
(Paresce \etal 1980), where $\partial^2 N / \partial N_\nHI\partial
z'$ is the redshift and column density distribution of absorbers along the line
of sight, and $\tau$ is the Lyman continuum optical depth through an individual
cloud. 

\subsection{Intervening Absorption}

The actual distribution of intervening clouds
is still quite uncertain, especially in the range $10^{16.3}\lta N_\nHI\lta
10^{17.3}\cmm$, where most of the contribution to the effective photoelectric
opacity actually occurs. As a function of \HI column, a single power-law with
slope $-1.5$  appears to provide a surprisingly good description over
nearly 10 decades in $N_\nHI$ (e.g. Hu \etal 1995).
At high redshifts, it is a good approximation to use for the distribution of 
absorbers: 
\begin{equation}
{{\partial^2N}\over{\partial N_\nHI\partial z}}=N_0\,N_\nHI^{-1.5}(1+z)^
{\gamma}. \label{eq:dis} 
\end{equation}
For simplicity, I will 
assume here a single redshift exponent, $\gamma=2$, for the entire range in
column densities. A normalization value of $N_0=4.0\times 10^7$ produces about
3 LLS per unit redshift at $z=3$, as observed by Stengler-Larrea \etal (1995),
and, at the same epoch, $\sim 150$ lines above
$N_\nHI=10^{13.77} \cmm$, in good agreement with the estimates of Kim \etal
(1997). With this normalization and $\gamma=2$, the adopted distribution
provides about the same \HI photoelectic opacity as in the model discussed by 
Haardt \& Madau (1996).

\subsection{Attenuation Length} 

If one extrapolates the $N_\nHI^{-1.5}$ power-law in equation (\ref{eq:dis}) to
very small and large columns, the effective optical depth becomes an analytical
function of redshift and wavelength, 
\begin{equation}
\tau_{\rm eff}(\nu_o,z_o,z)={4\over 3}\sqrt{\pi\sigma_0}\, N_0 \left({\nu_o
\over \nu_L}\right)^{-1.5} (1+z_o)^{1.5}\left[(1+z)^{1.5}-(1+z_o)^{1.5}\right], 
\label{eq:t} 
\end{equation}
where $\sigma_0$ is the hydrogen photoionization cross-section at the Lyman
edge $\nu_L$, and we have not included the contribution of helium to the
attenuation along the line of sight. 
It is practical, for the present discussion, to define a redshift $z_1(\nu_1)$
such that the effective optical depth between $z_o$ and $z_1$ is unity. When 
$\nu_o<\nu_L$, a photon emitted at $z_1$ with frequency $\nu_1=\nu_o (1+z_1)/
(1+z_o)$ will be redshifted below threshold before being significantly 
attenuated by intervening \HI. From equation (\ref{eq:t}), it can be shown that 
the universe will be optically thin below   
\begin{equation}
z_1(\nu_1)={1.616\,\nu_1/\nu_L\over [(\nu_1/\nu_L)^{3/2}-1]^{1/3}}-1.
\end{equation}
This expression has a minimum for $\nu_1=1.59\nu_L$, corresponding to what
we shall term in the following the ``breakthrough epoch'' of the universe,
$z_{\rm br}\equiv z_1(1.59\nu_L)=1.56$. It is only for $z<z_{\rm br}$ that 
the degree of ionization of the IGM will be 
determined by the balance between radiative recombinations and the 
total ionizing flux emitted by all QSOs which appear after $z_{\rm br}$. 

By contrast, due to the rapid increase in the number of intervening absorbers 
with lookback time, beyond a redshift of 2 the mean free 
path of photons at $912\,$\AA\ becomes so small that the radiation is 
largely ``local'', as sources at higher redshifts are severely absorbed. 
Expanding equation (\ref{eq:t}) around $z=3$, for example, one gets $\tau_{\rm 
eff} (\nu_L)\approx 0.36 (1+z)^2 \Delta z=1$ for $\Delta z=0.18$. This 
corresponds to a proper distance or ``absorption length'' of only 
\begin{equation}
\Delta l(\nu_L)\approx 33\, {\rm Mpc} \left({1+z\over 4}\right)^{-4.5}.
\end{equation}
In the local (or ``source-function'') solution to
the equation of radiative transfer, this strong attenuation effect is
approximated by simply ignoring sources with $z>z_1$ and neglecting absorption
for those sources with $z<z_1$.\footnote{As filtering through a clumpy IGM 
will steepen the UV background spectrum (Miralda-Escud\`e \& 
Ostriker 1990), the absorption length at the \HeII edge will be 
significantly smaller than at 1 ryd, $\Delta l(4\nu_L)\approx 1.5 \Delta l
(\nu_L) \sqrt{J_{228}/J_{912}}$.}~ Since only emitters within the
volume defined by an absorption length contribute significantly to the background 
intensity shortward of the
Lyman edge, cosmological effects such as source evolution and frequency shifts
can be neglected, and the number of ionizing photons per
unit proper volume at $z$ can be written as  
\begin{equation}
n_{\rm ion}\equiv {4\pi\over c} \int_{\nu_L}^\infty d\nu {J(\nu)\over h\nu}=
\dot n_{\rm ion} {\langle \Delta l \rangle\over c},
\end{equation}
where $\dot n_{\rm ion}(t)\equiv\int_{\nu_L}^\infty d\nu \epsilon(\nu,t)/h\nu$, 
and $\langle \Delta l \rangle$ is an average over the the incident photon 
spectrum.\footnote{Since $\Delta l(\nu)\propto (\nu/\nu_L)^{1.5}$, one
has $\langle \Delta l\rangle=\Delta l(\nu_L) \delta (4^{1.5-\alpha_s}-1)/
(1-4^{-\alpha_s})$, where $\delta\equiv \alpha_s/(1.5-\alpha_s)$,
$\epsilon(\nu)\propto \nu^{-\alpha_s}$, and we have assumed a cutoff at 4 ryd
because of \HeII absorption. A spectrum with $\alpha_s=1.8$ yields 
$\langle \Delta l\rangle=2.2\Delta l(\nu_L)$.}~ 
The small absorption length is mostly due to systems with continuum optical 
depth $\tau\sim 1$. Within the assumption that all absorbers are optically 
thin, highly ionized in both H and He, and contain most of the baryons of 
the universe, it is possible to derive a relation between the mean absorption 
length and the gas recombination time $t_{\rm rec}$ from the equation of 
ionization equilibrium, \begin{equation}
\langle \Delta l\rangle={n_{\rm ion}\over n_\nH}\, ct_{\rm rec}, \label{eq:dl} 
\end{equation}
where $n_\nH$ is the volume-averaged hydrogen density of the expanding IGM, 
$n_\nH(0)=1.3\times 10^{-7}$ $(\Omega_b h_{50}^2/0.06)$,
\begin{equation}
t_{\rm rec}=[(1+2\chi) n_\nH \alpha_B\,C]^{-1}=1.5\, {\rm Gyr} \left({\Omega_b
h_{50}^2 \over 0.06}\right)^{-1}\left({1+z\over 4}\right)^{-3} C_{10}^{-1}, 
\end{equation}
$\alpha_B$ is the recombination coefficient to the excited states of hydrogen,
$\chi$ the helium to hydrogen cosmic abundance ratio, $C\equiv \langle
n_\nHII^2\rangle/n_\nHII^2$ is the ionized hydrogen clumping 
factor,\footnote{This may be somewhat lower than the total gas clumping 
factor if higher density regions are less ionized (e.g. Gnedin \& Ostriker
1997).}\ and we assumed a gas temperature of $10^4\,$K. An empirical 
determination of the  clumpiness of the IGM at 
high redshifts is hampered by our poor knowledge of the ionizing background
intensity and the typical size and geometry of the absorbers. Numerical
N-body/hydrodynamics simulations of structure formation in the IGM within the
framework of CDM dominated cosmologies have recently provided a definite picture
for the origin of intervening absorption systems, one of an interconnected
network of sheets and filaments, with virialized systems located at 
their points of intersection.  In the simulations of Gnedin \& Ostriker (1997),
for example, 
the clumping factor rises above unity when the collapsed fraction of 
baryons becomes significant, i.e. $z\lta 20$, and grows to $C\gta 10$ (40) 
at $z\approx 8$ (5) (because of finite resolution effects, numerical
simulations will actually underestimate clumping). The recombination time is
then much shorter than that for a uniform IGM. 

\section{Reionization of the Universe}

In the rest of this talk I will show how 
a simple formalism can shed some light on the time-dependent reionization 
process, and assess the role of quasars and star-forming galaxies as candidate
sources of photoionization at high redshifts. 

\subsection{Time Evolution of \HII Filling Factor}

When an isolated point source of ionizing radiation turns on, the ionized
volume initially grows in size at a rate fixed by the emission of UV photons,
and an ionization front separating the \HII and \HI regions propagates
into the neutral gas. Most photons travel freely in the ionized bubble, and are
absorbed in a transition layer. Across the I-front the degree of
ionization changes sharply on a distance of the order of the mean free path of
an ionizing photon. When $t_{\rm rec}\ll t$, the growth of the \HII region is
slowed down by recombinations in the highly inhomogeneous IGM, and its evolution
can be decoupled from the expansion of the universe. Just like in the static
case, the ionized bubble will fill its time-varying Str\"omgren sphere
after a few recombination timescales.

The filling factor of \HII regions in the 
universe, $Q_\nHII$, is, at any given instant $t$, equal to the integral
over cosmic time of the rate of ionizing photons emitted per hydrogen atom and
unit cosmological volume by all radiation sources present at earlier epochs, 
\begin{equation}
\int_0^t {\dot n_{\rm ion}(t')\over n_\nH(t')}dt', 
\end{equation}
{\it minus} the rate of radiative recombinations,
\begin{equation}
\int_0^t {Q_\nHII(t')\over t_{\rm rec}(t')}dt'.
\end{equation}
Differentiating, one gets
\begin{equation}
{dQ_\nHII\over dt}={\dot n_{\rm ion}\over n_\nH}-{Q_\nHII\over t_{\rm rec}}.
\label{eq:qdot}
\end{equation}
It is this simple differential equation -- and its equivalent for
expanding helium zones -- that {\it statistically describes the transition
from a neutral universe to a fully ionized one}, independently, for a given
emissivity, of the complex and possibly short-lived emission histories of
individual radiation sources, e.g., on whether their comoving space density is
constant or actually varies with cosmic time (cf Arons \& Wingert 1972). In 
the case of a time-independent
clumping factor, equation (\ref{eq:qdot}) has formal solution 
\begin{equation}
Q_\nHII(t)=\int_0^t dt'\, {\dot n_{\rm ion}\over n_\nH}\,
\exp\left(-{t'\over t_{\rm rec}}+{t'^2\over t_{\rm rec}t }\right),
\end{equation}
with $t_{\rm rec}\propto t'^2$. At high 
redshifts, and for an IGM with $C\gg 1$, one can expand around $t$ to find 
\begin{equation}
Q_\nHII(t)\approx {\dot n_{\rm ion}\over n_\nH}t_{\rm rec}. 
\label{eq:qa}
\end{equation}
The porosity of ionized bubbles is approximately given then by the {\it number
of ionizing photons emitted per hydrogen atom in one recombination time}. In 
other words, because of hydrogen recombinations, only a fraction $t_{\rm
rec}/t$ ($\sim$ a few per cent at $z=5$) of the photons emitted above 1 ryd is
actually used to ionize new IGM material. The universe is completely reionized
when $Q=1$, i.e. when 
\begin{equation}
\dot n_{\rm ion} t_{\rm rec}=n_\nH.  \label{eq:Q1}
\end{equation}
Figure 1 shows the \HII (and \HeIII) filling factor as a function of redshift for 
a QSO photoionization model and a clumpy IGM with $C=1, 10,
20$ and 35. The transition from a neutral to a ionized universe takes place too 
late in this model, as late as $z=3.8$ for $C=35$, and never before $z=4.9$ even 
in the limiting case of a uniform medium. 

\begin{figure}
\epsfysize=8cm 
\hspace{3.5cm}\epsfbox{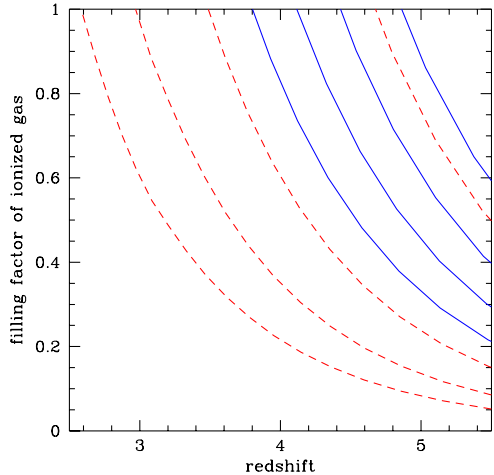} 
\caption[h]{The evolution of the \HII ({\it solid lines}) and \HeIII ({\it
dashed lines)} filling factors as a  function of redshift, for a universe where
photoionization is dominated by QSOs (from Haardt \etal 1998). The total 
IGM density is taken to be 
$\Omega_b h_{50}^2=0.06$, and, from right to left, the four curves assume
a clumping factor $C=1, 10, 20,$ and 35. The QSO intrinsic spectrum varies
as $\nu^{-1.8}$ shortward of the hydrogen Lyman edge. Note how the ionization
of \HI (\HeII) is never completed before $z=4.4$ ($z=3.5$) in models with 
$C\ge 10$.}
\end{figure}

\subsection{Critical Emissivity}

\begin{figure}
\epsfysize=8cm 
\hspace{3.5cm}\epsfbox{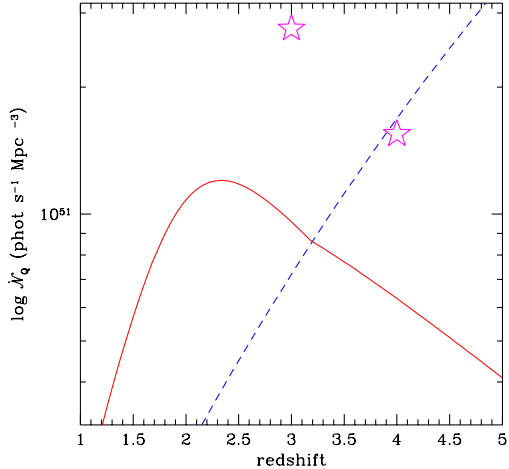}
\caption[h]{Comoving emission rate of hydrogen Lyman continuum photons ({\it solid
line}) from QSOs, compared with the minimum rate ({\it dashed line}) which is
needed to fully ionize an Einstein--de Sitter universe with $h_{50}=1$ and
$\Omega_b=0.06$ (from Haardt \etal 1998). Models based on photoionization 
by quasar sources fall short 
by about an order of magnitude at $z=5$. The {\it starred}
symbols show the estimated contribution of star-forming galaxies at $z=3$ and,
with larger uncertainties, at $z=4$. They should be considered as 
upper limits, as they have been derived assuming a negligible amount
of intrinsic \HI absorption.} 
\end{figure}

We can better quantify now the inadequacy of known sources of radiation to 
provide the required amount Lyman continuum photons at $z\approx 5$. Because of
the local character of the UV background at high redshifts, 
one can compute at any given epoch a critical value for the photon
emission rate per unit cosmological comoving volume, $\dot {\cal N}_{\rm ion}$,
independently of the (unknown) previous emission history of the universe: only
rates above  this value will provide enough UV photons to keep the IGM 
ionized at epoch. One can then compare our determinations of $\dot {\cal N}_{\rm
ion}$ to the estimated contribution from QSOs and star-forming galaxies. We
have seen in the previous sections that only photons emitted within one
recombination timescales or, equivalently, within one attenuation length of any
fluid element in the universe can actually be used in the equation of ionization
equilibrium. Equation (\ref{eq:Q1}) can then be rewritten as $\dot n_{\rm ion}
\langle \Delta l \rangle=n_\nH c$, or 
\begin{equation}
\dot {\cal N}_{\rm ion}={n_\nH(0)c\over 2.2 \Delta
l(\nu_L)}=(10^{48}\,\ndotunits)\, h_{50}(1+z)^{4.5} {\Omega_b h_{50}^2\over
0.06}, 
\end{equation}
where a spectral index of $\alpha_s=1.8$ has been assumed for the volume
emissivity between 1 and 4 ryd. The uncertainty on this critical rate is
difficult to estimate, as it depends on the statistics of intervening absorbers
and the nucleosynthesis constrained baryon density. A quick exploration of the 
available parameter space indicates that the uncertainty on $\dot 
{\cal N}_{\rm ion}$ is unlikely to exceed $\pm 0.3$ in the log. The evolution
of the critical rate as a function of redshift is plotted in Figure 2: 
$\dot {\cal N}_{\rm ion}$ is higher than the quasar contribution  already at 
$z\gta 3$. At $z=5$, the deficit of Lyman continuum photons is about an order 
of magnitude. For relatively bright galaxies to produce enough UV radiation at
$z=5$, their space density would have to be comparable to the one observed at
$z\approx 3$, with most ionizing photons being able to escape freely from the
regions of star formation into the IGM. This scenario appears quite improbable,
in light of the decrease in the UV galaxy emissivity observed in the HDF above
$z=2$ (Madau \etal 1996) and of direct observations of local starbursts below
the Lyman limit showing that at most a few percent of the stellar ionizing
radiation produced by these sources actually escapes into the IGM (Leitherer
\etal 1995). 

It is interesting to convert the derived value of $\dot {\cal N}_{\rm ion}$  
at $z=5$ into a star formation rate per unit (comoving) volume, $\dot \rho$:
\begin{equation}
{\dot \rho}(z=5)=\dot {\cal N}_{\rm ion} \times 10^{-53.1} f_{\rm esc}^{-1}
\approx 0.025 f_{\rm esc}^{-1}\, \sfrd, 
\end{equation}  
where $f_{esc}$ is the unknown fraction of Lyman continuum photons from
O stars which escapes the galaxy \HI layers into the IGM. 
The conversion factor assumes a Salpeter IMF with solar metallicity, and has
been computed using Bruzual \& Charlot (1998) population synthesis code. 
If dwarf galaxies associated with $\sigma\approx 20$ km s$^{-1}$ dark
matter halos are responsible for ionizing the universe at $z=5$,
and are able to convert a fraction $f\Omega_b$ of their total mass into 
stars over a timescale of order the free-fall time of the halo, it can be 
shown then that their comoving space density must be approximately
\begin{equation}
{0.025\, f_{esc}^{-1}\sfrd\over 0.2 f {\, \rm M_\odot yr^{-1}}}
\approx {0.13\over f_{esc} f} {\, \rm Mpc^{-3}},
\end{equation}
at least several hundred times larger than the space density of present-day
$L>L^*$ galaxies.


\end{document}